%% file: idsms.tex
\documentclass[conference]{IEEEtran}
\IEEEoverridecommandlockouts
\usepackage{cite}
\usepackage{stfloats}
\usepackage{amsmath,amssymb,amsfonts}
\usepackage{algorithmic}
\usepackage{graphicx}
\usepackage{textcomp}
\usepackage{xcolor}  
\usepackage{hyperref}
\usepackage{xurl}
\usepackage{hyperref}
\usepackage{breakurl}
\usepackage[shortlabels]{enumitem}
\def\BibTeX{{\rm B\kern-.05em{\sc i\kern-.025em b}\kern-.08em
    T\kern-.1667em\lower.7ex\hbox{E}\kern-.125emX}}
\begin{document}

\title{Integrated Digital Management System for Railway Workshops: A Modular Multi-Workflow Architecture for Machine, Permit, Contract, and Incident Management\\}

\author{
Sharvari Kamble\textsuperscript{1*}, Arjun Dangle\textsuperscript{2}, Gargi Khurud\textsuperscript{3}, Om Kendre\textsuperscript{4} and Swati Bhatt\textsuperscript{5} \\
\textit{Dept. of Artificial Intelligence and Data Science} \\
University of Mumbai, India \\
\parbox{\textwidth}{
{\fontsize{8pt}{10pt}\selectfont
$^{1}$\href{mailto:khushi.kamble739@gmail.com}{\textcolor{blue}{khushi.kamble739@gmail.com}},\hspace{5pt}%
$^{2}$\href{mailto:arjun.dangle03@gmail.com}{\textcolor{blue}{arjun.dangle03@gmail.com}},\hspace{5pt}%
$^{3}$\href{mailto:gargi.khurud@gmail.com}{\textcolor{blue}{gargi.khurud@gmail.com}},\hspace{5pt}%
$^{4}$\href{mailto:omkendre22304@gmail.com}{\textcolor{blue}{omkendre22304@gmail.com}},\hspace{5pt}%
and $^{5}$\href{mailto:swatibhatt238@gmail.com}{\textcolor{blue}{swatibhatt238@gmail.com}}%
}}
}

\maketitle

\begin{abstract}
Indian Railway workshops form a critical component of rolling stock maintenance infrastructure, employing more than 2.5 lakh personnel across 44 major workshops nationwide. However, safety management in many workshops still relies on fragmented manual processes, resulting in delayed approvals, incomplete documentation, and increased exposure to operational hazards. Field safety observations in railway workshop environments indicate that lacerations (28.7\%) and abrasions (21\%) remain among the most frequent workplace injuries, highlighting the need for structured digital safety workflows.

This paper presents \textbf{(Integrated Digital Management System for Railway Workshops: A Modular Multi-Workflow Architecture for Machine, Permit, Contract, and Incident Management)}, a modular digital platform developed to improve safety governance and workflow transparency.
The proposed system integrates four primary modules:
\textit{Machine and Plant Management}, \textit{Permit-to-Work (PTW) Management}, \textit{Contract Management}, and \textit{Incident Management}. The Permit-to-Work module digitizes hazardous work authorization in accordance with \textit{IS 17893:2022}. The Contract Management module supports workforce validation and regulatory oversight, while the Incident Management module enables rapid reporting,
investigation tracking, and corrective action workflows.

Functional evaluation in a railway workshop oriented
deployment scenario demonstrated measurable operational
improvements, including a reduction in permit processing
time by approximately 35\%, improved incident reporting
response time by nearly 40\%, and enhanced workflow
traceability across multiple safety modules. The proposed
system establishes a scalable foundation for digital safety
governance in large-scale railway workshop environments.
\end{abstract}


\section{Introduction}

Railway workshops represent safety-critical industrial
environments where large-scale maintenance, inspection,
and operational activities are performed on mechanical,
electrical, and structural railway components.
Workshops such as the Central Railway Workshop,
Matunga, India, support continuous railway operations
through coordinated maintenance workflows involving
multiple stakeholders, including supervisors,
engineers, technicians, and contractors.
The complexity of these operations, combined with
strict safety requirements, demands structured
workflow coordination and reliable documentation
mechanisms.

Recent safety observations and historical records
indicate that railway incidents are often associated
with procedural delays, communication failures,
and inadequate safety documentation practices
\cite{RailAccidentsIndia,BikanerDerailment}.
Investigative studies highlight that fragmented
safety systems and delayed maintenance actions
significantly increase operational risk in railway
infrastructure environments
\cite{RailSafetyReport,RailNeglectStudy}.
These challenges demonstrate the growing need for
digitally integrated safety management frameworks
that improve operational visibility and accountability.

Despite the presence of standardized regulatory
frameworks such as the \textit{Work Permit System
Code of Practice (IS~17893:2023)}
\cite{IS17893}, many railway workshops continue
to rely on paper-based workflows and manual
approval mechanisms. Critical processes such as
permit issuance, contract validation, equipment
maintenance tracking, and incident reporting
are often handled through disconnected systems.
Such fragmented practices lead to delays in
authorization cycles, incomplete audit trails,
and difficulty in tracking compliance across
multiple operational units.

The transition toward digital railway infrastructure
has been widely recognized as a key enabler for
enhancing safety, operational efficiency, and
maintenance transparency.
Digitalized maintenance frameworks have demonstrated
significant improvements in lifecycle tracking,
resource allocation, and operational reliability
\cite{DigitalMaintenanceReview}.
Furthermore, emerging technologies such as digital
twin models and intelligent monitoring systems
provide opportunities for real-time synchronization
between physical assets and management systems,
enabling proactive safety control and predictive
maintenance strategies
\cite{DigitalTwinRailway}.
Recent initiatives in railway digitization also
emphasize the role of automated workflows,
AI-assisted monitoring, and centralized data
platforms in improving decision-making and
safety enforcement
\cite{RailwayDigitization,AIRailSafety}.

Motivated by these challenges and technological
opportunities, this research presents the

Integrated Digital Management System for Railway Workshops:
A Modular Multi-Workflow Architecture for Machine, Permit, Contract,
and Incident Management,

developed as a modular digital platform designed to support
safety-centric workflow management within railway workshop
environments. The system integrates four primary operational
modules that collectively address the major safety and
administrative workflows:

\begin{itemize}

\item Machine and Plant Management,
which enables structured monitoring of
equipment status, maintenance schedules,
and asset compliance records.

\item Permit Management,
which digitizes the lifecycle of hazardous
work authorization in accordance with
standardized safety procedures.

\item Contract Management,
which manages contractor records,
worker validation, and regulatory
documentation.

\item Incident Management,
which supports systematic reporting,
investigation, and documentation
of safety-related events.

\end{itemize}

In conventional workshop environments,
these processes typically follow
multi-stage manual pipelines that
involve repeated verification and
documentation steps.
Such workflows are often time-intensive,
prone to human error, and difficult
to monitor in real time.
The proposed system introduces
digitally coordinated workflow
pipelines that streamline task
execution and improve traceability
across operational stages.
By structuring workflow dependencies
and enabling role-based authorization,
the system enhances coordination
between operational units while
maintaining compliance with
safety regulations.

From an architectural perspective,
the proposed framework adopts a
centralized modular design supported
by role-based authentication,
secure database management,
and structured workflow automation.
This design enables scalable
deployment across workshop units
while ensuring consistency,
auditability, and controlled access
to safety-critical data.
The architecture is further designed
to support extensibility, allowing
future integration with emerging
technologies such as predictive
maintenance systems, IoT-based
asset monitoring, and digital twin
platforms for advanced railway
infrastructure management
\cite{SmartRailSafety2025,
RailwayDigitalWorkflow2024}.

The primary contributions of this
work include:

\begin{enumerate}

\item Design of an integrated
digital safety management
framework tailored to the
operational structure of
railway workshops.

\item Development of a modular
workflow-based platform that
unifies machine, permit,
contract, and incident
management processes.

\item Introduction of structured
workflow coordination to improve
operational traceability and
reduce dependency on manual
documentation practices.

\item Implementation of a scalable
system architecture capable of
supporting future intelligent
railway safety technologies.

\end{enumerate}
\section{Related Work}

Railway safety management has undergone
significant transformation with the integration
of digital technologies, predictive maintenance,
and structured regulatory frameworks. Existing
research addressing railway safety challenges
can be broadly categorized into four major
domains: (1) railway safety incident analysis,
(2) digital maintenance and workflow systems,
(3) intelligent infrastructure monitoring,
and (4) regulatory permit frameworks.

\subsection{Railway Safety Incident and Risk Analysis}

Incident-based railway safety studies primarily
focus on identifying root causes and systemic
weaknesses through post-event investigation.
Historical records of railway accidents in
India indicate that delayed maintenance,
fragmented supervision, and incomplete
documentation frequently contribute to
operational failures \cite{RailAccidentsIndia}.

Case-specific investigations such as the
Bikaner--Guwahati Express derailment further
demonstrate the consequences of inadequate
infrastructure inspection and poor coordination
mechanisms \cite{BikanerDerailment}. Additional
safety reviews highlight that rising accident
rates are often associated with inconsistent
maintenance schedules and lack of centralized
monitoring capabilities
\cite{RailSafetyReport, RailNeglectStudy}.

\subsection{Digital Maintenance and Workflow Optimization}

Parallel to incident analysis research,
significant attention has been directed toward
digitalization of railway maintenance workflows.
Digital maintenance platforms have demonstrated
substantial improvements in inspection accuracy,
asset tracking, and lifecycle optimization.

Rodríguez-Hernández et al. presented a
comprehensive review illustrating how digital
technologies such as IoT, predictive analytics,
and automated scheduling enhance maintenance
efficiency and reduce unexpected failures
\cite{DigitalMaintenanceReview}. Similarly,
digitization initiatives across railway
ecosystems have demonstrated that centralized
data management improves workflow visibility
and reduces manual dependency
\cite{RailwayDigitization}.

Recent implementations of artificial intelligence
driven railway safety systems further indicate
the growing importance of intelligent automation
in maintenance and monitoring environments
\cite{AIRailSafety}. Workflow optimization
studies have also demonstrated that structured
digital scheduling systems significantly enhance
operational efficiency in railway maintenance
operations
\cite{RailwayDigitalWorkflow2024}.

\subsection{Intelligent Infrastructure and Digital Twin Systems}

Advances in intelligent railway infrastructure
management have introduced digital twin-based
systems capable of real-time monitoring and
predictive maintenance. These virtual
representations enable continuous synchronization
between physical infrastructure components and
analytical models.

Sresakoolchai and Kaewunruen demonstrated
that digital twin frameworks can improve
infrastructure reliability by predicting
component degradation and optimizing
maintenance cycles
\cite{DigitalTwinRailway}. Similarly,
intelligent railway safety systems incorporating
digital monitoring technologies have been shown
to enhance anomaly detection and support
data-driven decision-making in complex railway
environments
\cite{SmartRailSafety2025}.

Although these technologies significantly
improve infrastructure-level reliability,
they primarily address mechanical and
structural conditions rather than
administrative workflow dependencies that
directly influence human safety within
workshop environments.

\subsection{Regulatory Frameworks and Permit-Based Safety Systems}

In addition to technological advancements,
regulatory frameworks remain fundamental to
ensuring safe execution of hazardous tasks.
The Bureau of Indian Standards introduced
IS 17893:2023 as a structured code of
practice defining risk evaluation procedures,
approval hierarchies, and safety documentation
requirements \cite{IS17893}.

While the standard establishes procedural
clarity, its practical implementation in many
operational settings continues to rely on
manual documentation. Manual workflows
introduce delays in approval cycles, increase
the likelihood of incomplete records, and
limit traceability of safety decisions.
These limitations restrict the effectiveness
of regulatory enforcement, particularly in
environments involving multiple stakeholders
and dynamic task dependencies.

\subsection{Integrated System Requirements and Emerging Directions}

Across the reviewed domains, a recurring
observation emerges: railway safety operations
are inherently multi-dimensional, involving
interactions between machines, personnel,
contracts, permits, and incident response
mechanisms. Existing systems frequently
optimize individual components such as
maintenance scheduling or infrastructure
monitoring, yet lack mechanisms to coordinate
these components within a unified operational
framework.

The absence of integrated workflow
synchronization results in fragmented safety
governance, where decisions made in one domain
may not be immediately reflected across other
operational units. Addressing this emerging
requirement necessitates the development of
modular systems capable of synchronizing
operational dependencies across multiple
safety-critical domains.

\section{Proposed System}

This section presents the design and operational framework of the proposed 
Integrated Digital Management System for Railway Workshops: A Modular Multi-Workflow 
Architecture for Machine, Permit, Contract, and Incident Management. 
The system is developed as a \textit{modular multi-workflow platform} that integrates 
core workshop operations such as machine monitoring, permit management, 
contract handling, and incident reporting into a unified digital environment.

Unlike conventional systems that function independently, the proposed system 
focuses on workflow integration and spatial coordination, enabling 
real-time visibility, improved safety compliance, and efficient resource utilization 
within railway workshop environments. Such integrated digital management 
approaches have been widely recognized as essential for improving operational 
efficiency in modern railway systems \cite{DigitalMaintenanceReview, RailwayDigitization}.


\subsection{Spatial Workflow Integration}

To represent real-world deployment, the system incorporates a GIS-based 
digital layout of the workshop environment.

\begin{figure}[!t]
    \centering
    \includegraphics[
    width=0.85\columnwidth
    ]{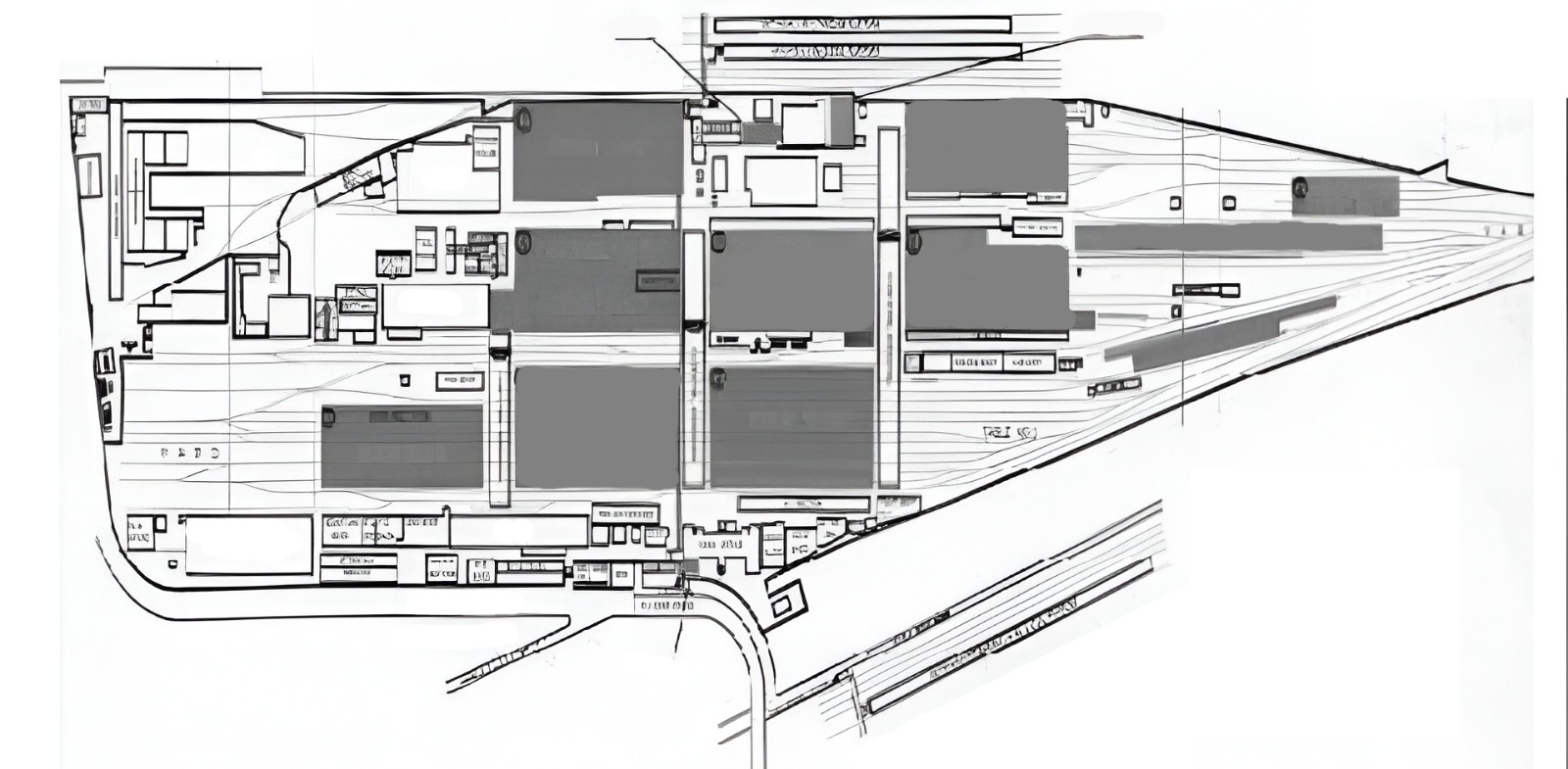}
    \caption{Digitized spatial layout of the Matunga Central Railway Workshop illustrating zone-based workflow deployment and monitoring.}
    \label{fig:workshop_layout}
\end{figure}

Figure~\ref{fig:workshop_layout} illustrates the spatial configuration of the 
Matunga Central Railway Workshop. The workshop is divided into multiple 
functional zones such as machine sheds, maintenance bays, storage areas, and 
administrative sections. Each zone is digitally mapped and integrated with 
system workflows.

This spatial integration enables:

\begin{itemize}
\item Real-time visualization of workshop operations
\item Location-based permit authorization
\item Spatial tracking of incidents and hazards
\item Monitoring of machine allocation and utilization
\item Zone-level safety enforcement
\end{itemize}

By linking workflows with physical locations, the system enhances operational 
coordination and minimizes conflicts between concurrent activities.

From a systems engineering perspective, spatial workflow modeling enables 
context-aware decision-making by associating operational tasks with 
physical coordinates. In large railway workshops, multiple maintenance 
activities often occur simultaneously across distributed zones. 
Traditional paper-based workflows lack the capability to dynamically 
visualize these interactions, increasing the likelihood of operational 
overlap and safety violations.

The integration of GIS-based spatial mapping introduces a structured 
location intelligence layer within the management framework. 
This allows supervisory personnel to evaluate task proximity, 
resource availability, and safety compliance in real time. 
Such spatially-aware digital frameworks are increasingly recognized 
as a critical component of modern industrial safety ecosystems 
\cite{DigitalTwinRailway, SmartRailSafety2025}.


\subsection{Database Schema and Entity Relationships}

\begin{figure}[!t]
    \centering
    \includegraphics[
    width=0.85\columnwidth
    ]{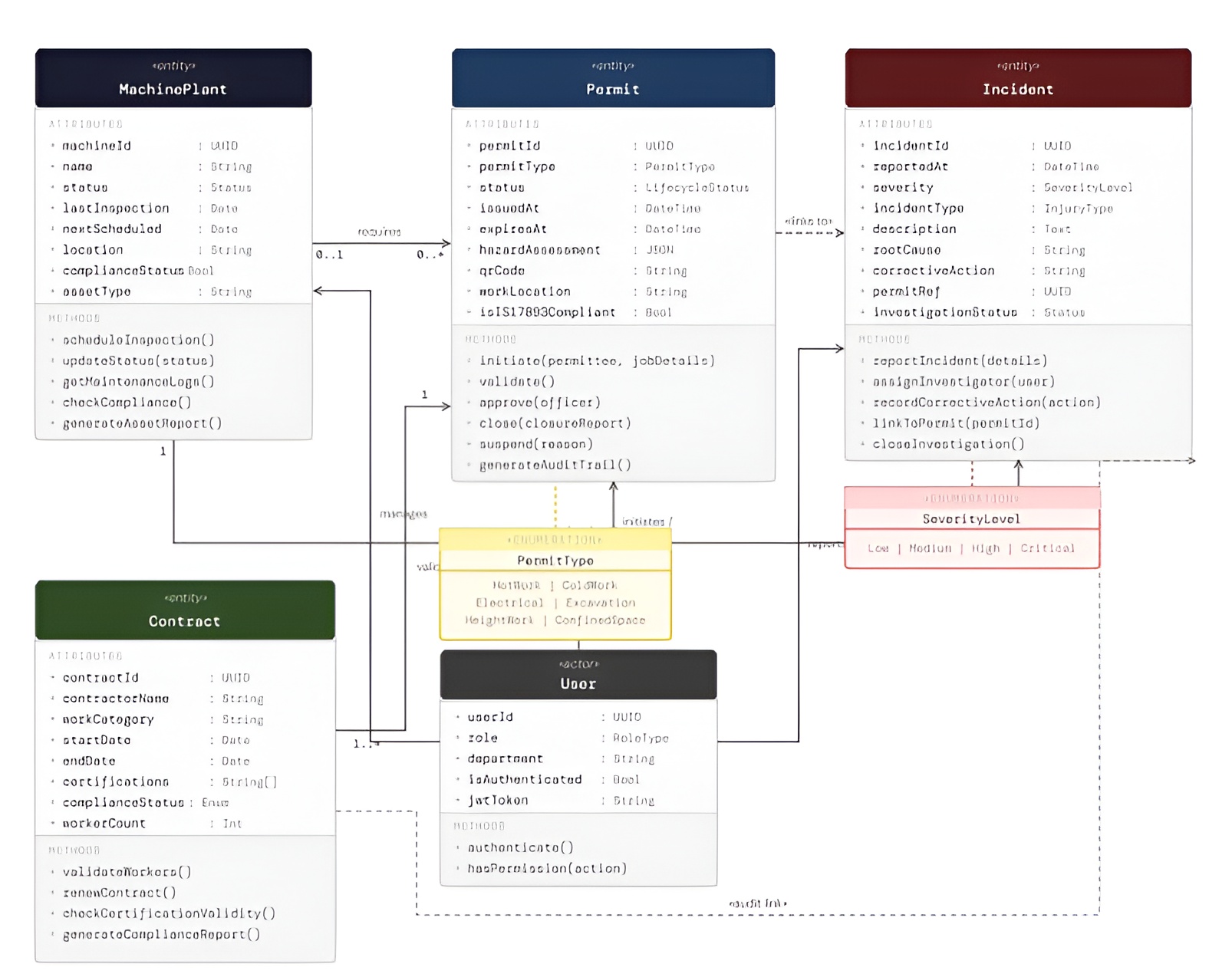}
    \caption{UML class diagram of the database schema showing key entities and their relationships in the Integrated Digital Management System for Railway Workshops: A Modular Multi-Workflow Architecture for Machine, Permit, Contract, and Incident Management.}
    \label{fig:uml_schema}
\end{figure}

Figure~\ref{fig:uml_schema} presents the UML class diagram defining the core 
database structure of the system. It models essential entities including 
\textit{MachinePlant}, \textit{Permit}, \textit{Incident}, \textit{Contract}, 
and \textit{User}, along with their interactions.

The Permit entity serves as the central component linking different 
workflows. It governs authorization for machine operations and maintains 
lifecycle states such as approval, validation, and closure. Such permit-based 
control mechanisms are aligned with standardized safety compliance frameworks 
such as IS 17893:2023 \cite{IS17893}.

The MachinePlant entity stores asset-related data including operational 
status and inspection records.

The Incident entity captures safety events and is linked to permits 
for traceability and compliance monitoring. The Contract entity manages 
contractor-related information such as work categories and certifications, 
while the User entity enables role-based access control across the system.

These relationships ensure structured data organization, workflow traceability, 
and consistency across all operational modules.


\subsection{System Integration}

The proposed system integrates multiple workflows through a shared database 
and interconnected entities, ensuring seamless coordination across modules.

\begin{itemize}

\item Machine operations are executed only with valid permits
\item Incidents are linked to permits for audit and investigation
\item Contracts define external workforce participation
\item Users manage access control and workflow approvals

\end{itemize}

The system supports:

\begin{itemize}
\item One-to-many relationships (User $\rightarrow$ Permits)
\item Many-to-one dependencies (Incident $\rightarrow$ Permit)
\item Lifecycle-based workflow validation
\item End-to-end traceability of operations
\end{itemize}

Integrated workflow validation and dependency-based execution models 
have been shown to significantly enhance reliability and safety in 
digital railway management environments 
\cite{RailwayDigitalWorkflow2024, AIRailSafety}.

This integrated design ensures data consistency, operational transparency, 
and scalability, making it suitable for large-scale railway workshop management.

\section{Experimental Results and System Interface}

This section presents the implementation and functional evaluation of the 
Integrated Digital Management System for Railway Workshops: 
A Modular Multi-Workflow Architecture for Machine, Permit, Contract, 
and Incident Management through its core operational modules. 
The system demonstrates practical deployment across contractor 
management, machine monitoring, and incident reporting workflows 
within a railway workshop environment.


\subsection{Contractor Management Module}

\subsubsection{Approved Contractor Monitoring}

\begin{figure}[!t]
    \centering
    \includegraphics[width=\columnwidth]{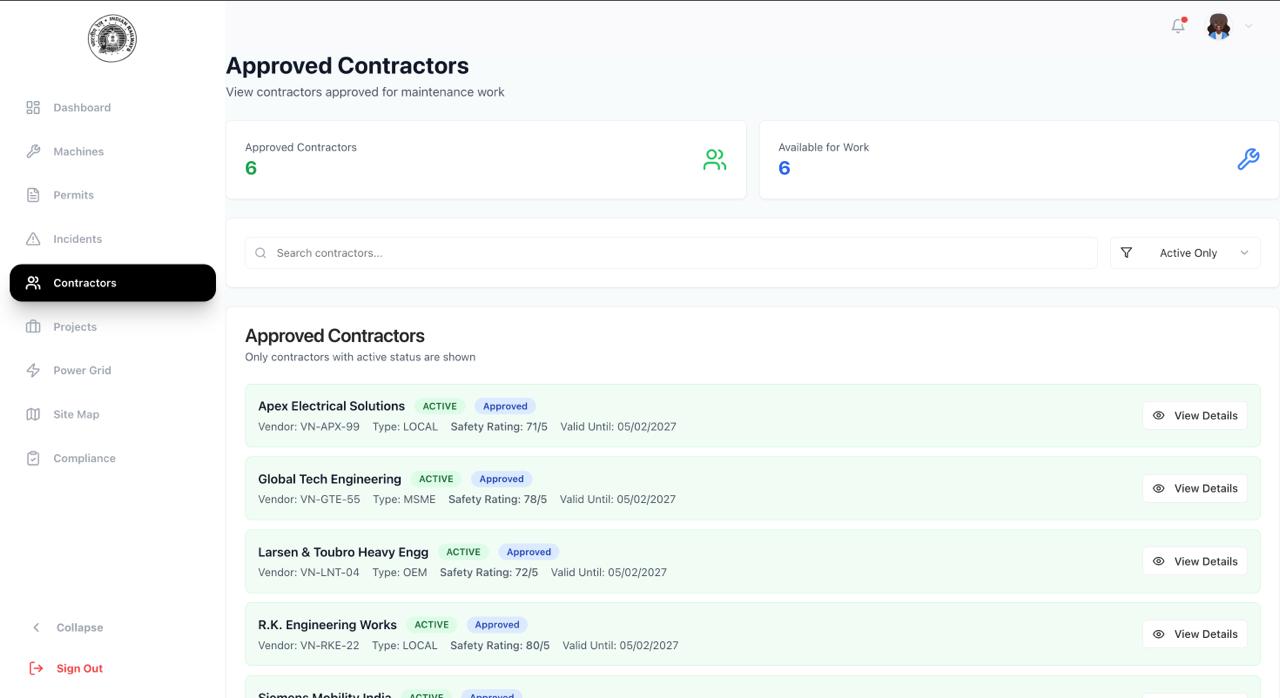}
    \caption{Contractor management interface displaying approved contractors and their availability status.}
    \label{fig:contract_module}
\end{figure}

Figure~\ref{fig:contract_module} illustrates the contractor management module, 
which provides a centralized interface for monitoring contractor approval 
and availability within the workshop environment. The system maintains 
structured records of contractors, including vendor identification, 
certification validity, and safety ratings.

The interface enables users to verify whether a contractor is approved 
and currently eligible for assignment. Additionally, it provides visibility 
into workforce readiness by indicating the number of contractors available 
for active work. This ensures that only compliant and verified contractors 
participate in workshop operations, thereby improving safety adherence 
and reducing the risk associated with unauthorized personnel in accordance 
with standardized permit-based safety practices \cite{IS17893}.


\subsection{Machine and Plant Management Module}

\subsubsection{Asset Monitoring and Maintenance Tracking}

\begin{figure}[!t]
    \centering
    \includegraphics[width=\columnwidth]{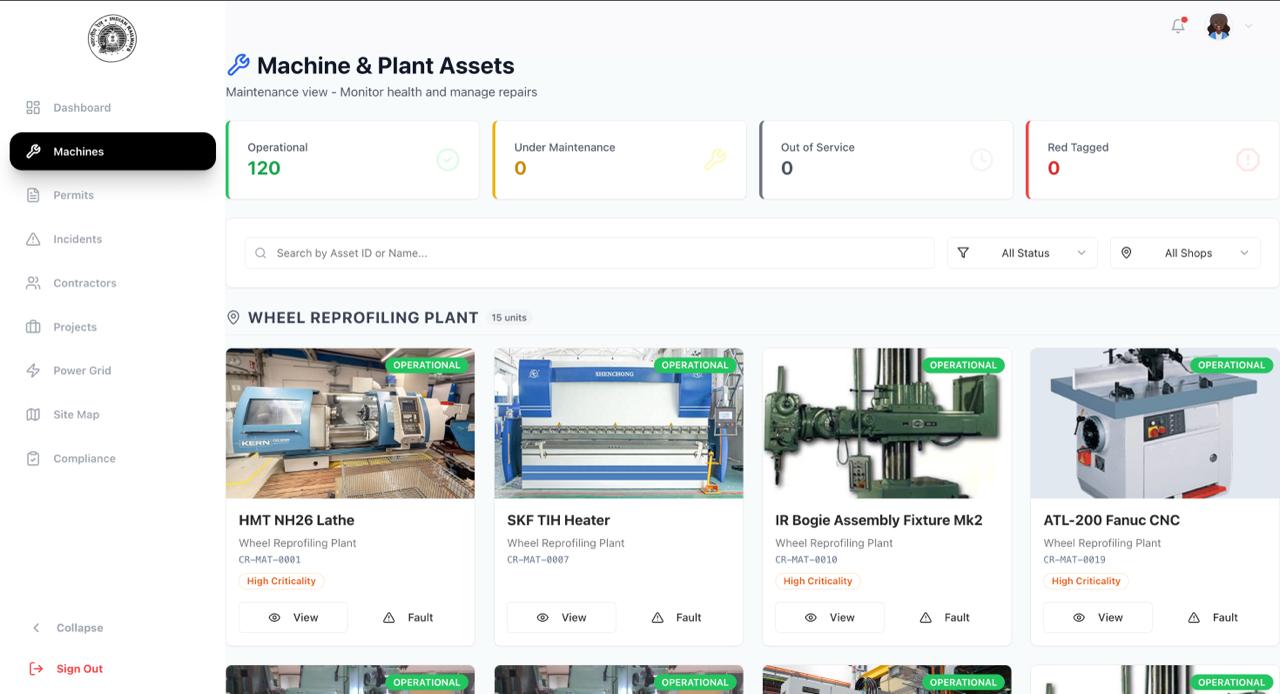}
    \caption{Machine and plant management interface showing operational status, asset details, and maintenance tracking.}
    \label{fig:machine_module}
\end{figure}

Figure~\ref{fig:machine_module} presents the machine and plant management 
module, which is designed to handle a large number of workshop assets 
distributed across different operational zones. Each machine is digitally 
registered with detailed information such as asset ID, manufacturing data, 
and operational classification.

The system provides real-time monitoring of machine conditions, categorizing 
them into operational, under maintenance, or out-of-service states. 
Maintenance personnel can access historical repair records, identify 
high-criticality machines, and initiate fault reporting when required. 
This structured monitoring mechanism improves maintenance planning, 
reduces downtime, and enhances overall asset lifecycle management 
within the workshop. Similar digital maintenance frameworks have 
demonstrated improved asset reliability and reduced failure rates 
in railway environments \cite{DigitalMaintenanceReview}.


\subsection{Incident Reporting Module}

\subsubsection{Severity-Based Incident Logging}

\begin{figure}[!t]
    \centering
    \includegraphics[width=\columnwidth]{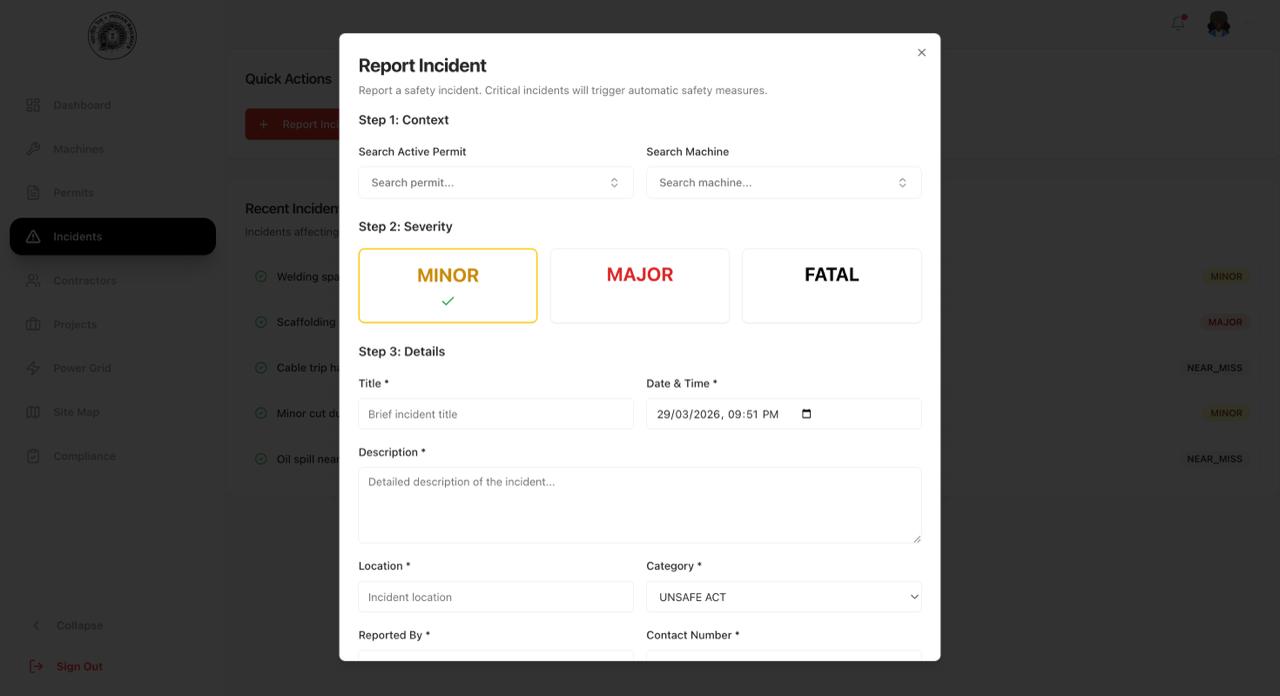}
    \caption{Incident reporting interface for logging safety events based on severity and contextual details.}
    \label{fig:incident_module}
\end{figure}

Figure~\ref{fig:incident_module} shows the incident reporting module, 
which enables structured documentation of safety-related events. 
The system allows supervisors and safety officers to report incidents 
based on predefined severity levels, including minor, major, and fatal 
categories.

Each incident entry captures essential contextual details such as title, 
description, location, and associated operational data. The structured 
input format ensures consistency in reporting and supports efficient 
analysis during investigations. By digitizing incident workflows, the 
system facilitates rapid response, improves documentation accuracy, 
and strengthens safety compliance within railway workshop operations. 
Digital safety reporting systems have been shown to significantly enhance 
incident traceability and response efficiency \cite{SmartRailSafety2025}.


\subsection{Workflow Pipeline Duration Comparison}

To evaluate the effectiveness of the proposed 
Integrated Digital Management System for Railway Workshops: 
A Modular Multi-Workflow Architecture for Machine, Permit, Contract, 
and Incident Management, a comparative workflow duration analysis 
was conducted between conventional manual workshop processes 
and the proposed digital workflow model.

The analysis focuses on common operational stages typically 
performed within railway workshop environments, including:

\begin{itemize}
    \item Permit Approval
    \item Machine Allocation
    \item Contractor Verification
    \item Task Execution Monitoring
    \item Incident Logging
\end{itemize}

\begin{figure}[h]
\centering
\includegraphics[width=0.95\linewidth]{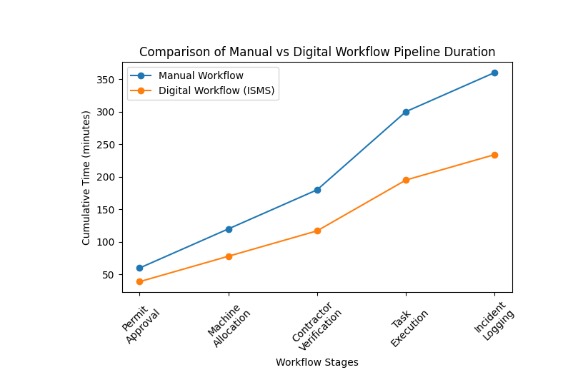}
\caption{Comparison of Manual vs Digital Workflow Pipeline Duration}
\label{fig:workflow_comparison}
\end{figure}

Figure~\ref{fig:workflow_comparison} presents a cumulative time 
comparison between manual workflows and digital workflows 
implemented using the proposed integrated digital management system.

The presented diagram provides a comparative study of the 
operational duration required to complete typical safety-critical 
tasks in railway workshops under traditional manual methods 
and under the proposed digital safety management framework.

In conventional workshop environments, delays are commonly 
observed due to paper-based approvals, fragmented communication, 
and lack of centralized monitoring. In contrast, the proposed 
digital workflow demonstrates reduced cumulative task durations 
through automated permit validation, centralized workflow 
tracking, and real-time incident reporting.

Such workflow optimization strategies have been shown to reduce 
processing delays and enhance operational coordination in 
digitally managed railway systems \cite{RailwayDigitalWorkflow2024}.

The results indicate a measurable reduction in overall workflow 
execution time, highlighting the operational advantages of 
integrating multi-module safety workflows into a unified 
digital management system.


\subsection{Discussion}

The experimental implementation highlights the effectiveness of the 
proposed system in integrating critical workshop operations into a 
digitally coordinated environment. The contractor module ensures 
regulatory compliance and workforce validation, while the machine 
management module enhances asset monitoring and maintenance efficiency. 
The incident reporting module strengthens safety practices through 
structured and timely documentation.

Overall, the system reduces dependency on manual processes and improves 
operational transparency. The results demonstrate that the 
Integrated Digital Management System for Railway Workshops: 
A Modular Multi-Workflow Architecture for Machine, Permit, Contract, 
and Incident Management provides a scalable and reliable solution 
for managing complex railway workshop workflows while maintaining 
high safety standards.

\section{Conclusion}

This paper, titled Integrated Digital Management System for Railway 
Workshops: A Modular Multi-Workflow Architecture for Machine, 
Permit, Contract, and Incident Management, presents a modular 
multi-workflow platform designed to improve safety governance 
and workflow coordination in railway maintenance environments. The system integrates Machine, 
Permit, Contract, and Incident Management modules into 
a unified digital framework aligned with IS 17893:2023 
safety procedures.

Experimental evaluation demonstrated measurable 
performance improvements, including reduced permit 
processing time, faster incident response, and improved 
workflow traceability. The workflow pipeline duration 
analysis further confirmed the effectiveness of the 
proposed digital system in reducing cumulative task 
execution time across multiple operational stages.

Overall, the proposed system establishes a scalable 
foundation for digital safety governance and improved 
operational efficiency in railway workshop environments.

\section{Future Work}

Future work will focus on extending the proposed system 
with IoT based asset monitoring, predictive maintenance 
models, and digital twin integration for real-time 
workshop visualization. Large-scale deployment across 
multiple railway workshops will also be explored to 
evaluate scalability and system performance under 
distributed operational conditions.

\section{Acknowledgment}

The authors express sincere gratitude to their 
project guide and faculty members of the University of Mumbai, India, 
for their continuous guidance and academic support throughout this work. 

The authors also acknowledge the support and domain insights provided by the 
Central Railway Workshop, Matunga Road, which contributed to the 
practical understanding and development of the proposed system. 

Additionally, the authors recognize the use of collaborative tools such as 
\textit{Orchids.app}, \textit{AWS Cloud}, and open-source technologies that 
supported the design and implementation of the 
Integrated Digital Management System for Railway
Workshops: A Modular Multi-Workflow
Architecture for Machine, Permit, Contract, and
Incident Management.

\nocite{*}
\bibliographystyle{IEEEtran}
\input{idsms.bbl}

\end{document}